\documentclass{optica-article}
\journal{opticajournal} 
\articletype{Research Article}
\usepackage{lineno}

\begin{document}
\title{Quantum surface effects on quantum emitters coupled to surface plasmon polariton}
\author{Xin-Yue Liu,\authormark{1,2} Chun-Jie Yang,\authormark{3} and Jun-Hong An\authormark{1,2,*}}
\address{\authormark{1}School of Physical Science and Technology \& Lanzhou Center for Theoretical Physics, Lanzhou University, Lanzhou 730000, China\\
\authormark{2}Key Laboratory of Quantum Theory and Applications of MoE \& Key Laboratory of Theoretical Physics of Gansu Province, Lanzhou University, Lanzhou 730000, China\\
\authormark{3}School of Physics, Henan Normal University, Xinxiang 453007, China}
\email{\authormark{*}anjhong@lzu.edu.cn} 


\begin{abstract*}
As an ideal platform for exploring strong quantized light-matter interactions, surface plasmon polariton (SPP) has inspired many applications in quantum technologies. Recent experiments discovered that quantum surface effects (QSEs) of the metal, including nonlocal optical response, electron spill-out, and Landau damping, invalidate the classical electromagnetic theory and contribute additional loss sources to the SPP in the nanoscale. This hinders its applications. Going beyond the widely used classical local response approximation, we use the Feibelman $d$-parameter method to investigate the QSE-modified non-Markovian dynamics of quantum emitters (QEs) coupled to a SPP in a planar metal-dielectric nanostructure. A mechanism to overcome the dissipation of the QEs caused by the lossy SPP with the QSEs is discovered. We find that, as long as the QE-SPP bound states are formed, a dissipationless entanglement among the far-separated QEs is created. Compared with the local-response approximate results, the QSEs play a constructive role in establishing such a coherent correlation. The result lays a foundation for understanding the light-matter interactions in absorptive media and paves the way for the application of SPP in quantum network.
\end{abstract*}

\section{Introduction}
Confining light in scales below the diffraction limit, surface plasmon polariton (SPP) has attracted extensive attention to explore strong light-matter interactions \cite{Tame2013,T_2015,Hu2024,Mueller2020,PhysRevLett.124.063902,PhysRevLett.128.167402,Yang2022} and led to fascinating applications in quantum technologies \cite{Chang2007,Dhama2022,Ho2016,Lee2021,doi:10.1126/sciadv.abn2026,Kongsuwan2019,Gao2021}. Recent advances in nanotechnologies have enabled the precise fabrication of ultra-fine plasmonic nanostructures, in which the characteristic size of the system is reduced to the nanoscale \cite{Baumberg2019,Scholl2012,Chikkaraddy2016,Sigle2015,doi:10.1021/acsnano.9b01651,Maniyara2019}. Interestingly, it was found that the experimental observations on the SPP in this scale exhibit considerable discrepancies from the theoretical predictions based on classical macroscopic electrodynamics under the local response approximation (LRA) \cite{Boroviks2022,Khalid2020,PhysRevLett.129.197401,Ciraci2019,Savage2012,Goncalves2021}. Hence, understanding the plasmon-mediated light-matter interactions beyond the LRA is a key problem in quantum plasmonics \cite{PhysRevB.104.L201405,PhysRevLett.118.157402,Goncalves2020}. 

\begin{figure}[tbp]
\centering
\includegraphics[width=0.8\columnwidth]{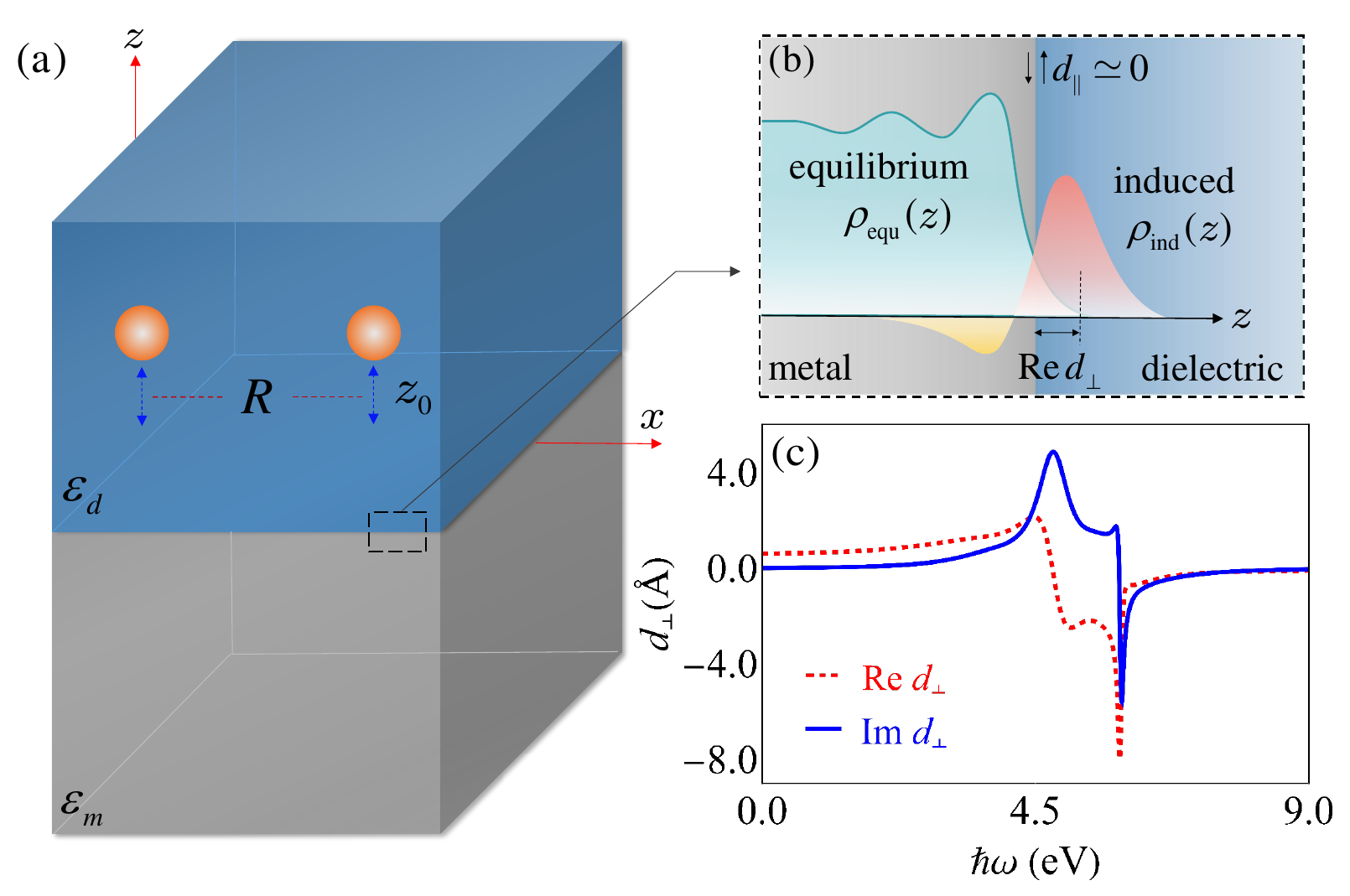}
\caption{(a) Schematic diagram of QEs separated by $R$ and positioned at $z_0$ above the metal. $\varepsilon_d$ is the permittivity of the dielectric and $\varepsilon_m$ is dielectric function of the metal. (b) A zoom-in view of the interface illustrates the equilibrium and induced electron densities and the Feibelman $d$-parameters. (c) $d_{\perp}$ of sodium by fitting the data extracted from ab initio calculations. $d_{\|}$ vanishes due to the charge-neutrality.}\label{Fig1}
\end{figure}

Including nonlocal optical response, electron spill-out, and Landau damping \cite{Raza2015,PhysRevLett.115.193901,Raza_2015,PhysRevX.11.011049,doi:10.1021/acs.jpcc.4c07319}, the quantum surface effects (QSEs) of the metal-dielectric interface become so strong that the plasmon-mediated light-matter interactions are dramatically modified when the system sizes reduce to the mean free path of electrons \cite{Yang2019,Raza2013,Stamatopoulou:22,Zhu2016}. Microscopic treatment of plasmon excitations in a quantum mechanical setting is enabled by time-dependent density functional theory (TDDFT) \cite{Marques2004}. However, this method is typically limited by its complexity and is applicable only to small plasmonic systems. To overcome this limitation, a nonclassical theory called the Feibelman $d$-parameter method has been proposed to treat the QSEs. This method involves the introduction of the induced surface charge and current, which are incorporated into two surface response parameters via the modified mesoscopic boundary conditions for the Maxwell equations \cite{FEIBELMAN1982287,PhysRevLett.118.157402,Yang2019,Goncalves2020,Babaze2023}. It bridges the gap between microscopic and macroscopic descriptions. It has been demonstrated that the QSEs play a significant role in a wide range of phenomena involving plasmon-mediated light-matter interactions, i.e., the enhancement of the Purcell factor \cite{PhysRevB.105.125419,Eriksen2024}, the electron energy-loss spectroscopy \cite{gonccalves2023interrogating}, the dipole-forbidden transitions and the two-photon emission \cite{Goncalves2020}, extinction cross-section \cite{Eremin2025}. A striking feature of QSEs is that they induce an ancillary metal absorption to the SPP, making the lifetime of the SPP even shorter than the one predicted by the LRA \cite{PhysRevB.104.L201405,PhysRevB.96.085413}. This severely restricts the applications of the SPP as a quantum bus in long-distance quantum devices \cite{PRXQuantum.2.017002,PhysRevA.109.033518}. The suppression of the destructive influence of lossy SPP on QEs is highly desired in the application of the SPP in quantum technologies.

Here, we thoroughly investigate the differences of the LRA and the QSEs for the strong-coupling dynamics between $N$ far-separated QEs and a common SPP in a planar metal-dielectric nanostructure, based on the macroscopic quantum electrodynamics (QED) and the Feibelman $d$-parameter method. A mechanism to overcome the dissipation of the QEs induced by the enhanced lossy SPP by the QSEs is uncovered. It is revealed that the dissipation of the QEs is suppressed and even a coherent quantum correlation among the QEs is persistently established whenever the bound states are formed in the energy spectrum of the total QE-SPP system. It makes the quantum network with QEs as nodes interconnected by the SPP realizable. Solving the dissipation problem of the QEs induced by the lossy SPP in the presence of the QSEs, our finding enriches our understanding of light-matter interactions in an absorptive medium at the nanoscale and paves the way for utilizing the SPP in long-distance quantum devices.

The paper is organized as follows. The system of $N$ QEs interacting with the SPP on a planar metal-dielectric interface in the presence of the QSEs is given in Sec. \ref{part1}. The exact non-Markovian dynamics and the constructive role played by the QE-SPP bound states in suppressing the dissipation of the QEs induced by the QSEs enhanced lossy SPP are analytically derived in Sec. \ref{part2}. The numerical results are given in Sec. \ref{numr}. A discussion on the experimental realizability and a summary are made in Sec. \ref{conclusions}. The Feibelman method for treating the QSEs is introduced in Appendix \ref{Appen-sec1} for self-consistency. The formulation of the Green's tensor of the SPP on the planar metal-dielectric structure is collected in Appendix \ref{Appen-Green}. 

\section{System}\label{part1}
We study the interactions between $N$ QEs and the SPP triggered by the radiation fields of the QEs on a planar metal-dielectric interface, see Fig. \ref{Fig1}(a). The QEs separated in a distance $R$ are embedded in the dielectric at a distance $z_0$ above the interface and are modeled by two-level systems with a common frequency $\omega_0$. Their radiation field triggers three modes, i.e., the radiative mode in the dielectric, the non-radiative mode absorbed by the metal, and the SPP mode hybridized by the field and the electron-density wave on the metal-dielectric interface. The distinguished character of the SPP is that it confines the electromagnetic field (EMF) on the metal-dielectric interface beyond the diffraction limit. It makes the system an ideal platform to realize the strong light-matter interactions. To describe the quantum features in these interactions, a rigorous quantization scheme incorporating the three modes in inhomogeneous absorptive medium is needed. A macroscopic QED method based on the dyadic Green's tensor has been developed via describing the metal absorption to the EMF by a quantum noise, which guarantees the canonical commutation relations of the quantized field \cite{PhysRevA.53.1818,Rivera2020,PhysRevLett.127.013602,PhysRevLett.122.213901,PhysRevLett.128.167403,Vazquez-Lozano2023,doi:10.1515/nanoph-2020-0451}. The electric dipole interaction Hamiltonian of the total system reads
\begin{eqnarray}\label{Hami}
\hat{H}=\sum_{i=1}^{N}\hbar \omega _{0}\hat{\sigma}_{i}^{\dag }\hat{\sigma}_{i}+\int_{0}^{\infty }\mathrm{d}\omega \Big\{ \int \mathrm{d}^{3}\mathbf{r}\hbar \omega \mathbf{\hat{
f}}^{\dag }(\mathbf{r},\omega )\cdot \mathbf{\hat{f}(r},\omega ) -\sum_{i=1}^{N}\lbrack \pmb{\mu}_{i}\cdot \mathbf{\hat{E}}(\mathbf{r}_{i},\omega )\hat{\sigma}_{i}^{\dag }+\text{H.c.}]\Big\},
\end{eqnarray}
where $\hat{\sigma}_{i}=|g_{i}\rangle \langle e_{i}|$ is the transition operators from the excited state $|e_{i }\rangle $ to the ground state $|g_{i}\rangle$ and $\pmb{\mu}_{i}$ is the dipole moment of the $i$th QE. The quantized electric field reads $\mathbf{\hat{E}}(\mathbf{r},\omega )=i\sqrt{\hbar \omega ^{4}/\pi\varepsilon _{0}c^{4}}\int \mathrm{d}^{3}\mathbf{r}^{\prime }\sqrt{\text{Im}[\varepsilon _{m}(\mathbf{r},\mathbf{r^{\prime }},\omega )]}\mathbf{G(r},\mathbf{r}^{\prime },\omega )\cdot \mathbf{\hat{f}}(\mathbf{r}^{\prime},\omega )$, where $\varepsilon _{0}$ is the vacuum permittivity, $\varepsilon _{m}({\bf r},{\bf r'},\omega )$ is the nonlocal dielectric function of the metal, $c$ is the speed of light, and $\mathbf{\hat{f}}(\mathbf{r},\omega )$ fulfilling $[\mathbf{\hat{f}(r},\omega ),\mathbf{\hat{f}}^{\dag }(\mathbf{r}^{\prime},\omega ^{\prime })]=\delta (\mathbf{r}-\mathbf{r}^{\prime })\delta (\omega
-\omega ^{\prime })$ is the annihilation operator of the field. The Green's tensor $\mathbf{G(r},\mathbf{r}^{\prime },\omega )$, denoting the field in frequency $\omega $ at $\mathbf{r}$ triggered by a point source at $\mathbf{r}^{\prime }$, is the solution of the Maxwell-Helmholtz equation $[{\pmb\nabla} \times {\pmb\nabla} \times -\omega^{2}c^{-2}\varepsilon _{m}\left({\bf r},{\bf r}', \omega \right) ]\mathbf{G(r},\mathbf{r}^{\prime },\omega )=\mathbf{I}\delta (\mathbf{r}-\mathbf{r}^{\prime })$, where $\mathbf{I}$ is the identity matrix. The scheme collects all the impacts of the metal-dielectric structure in the Green's tensor.

The surface response of metal to the EMF was traditionally described under the LRA, in which the response is nothing other than an induced surface charge. However, the classical treatment is insufficient when the QE-interface distance reduces to the mean free path of electrons \cite{Yang2019,Raza2013,Stamatopoulou:22}. The main missing elements are collected in the QSEs, which have three consequences. The first is the spatial nonlocality of the dielectric function of the metal \cite{PhysRevB.97.245405,Raza2015,Tserkezis2016a}. The second is the spill-out of electrons from the metal \cite{Toscano2015,PhysRevB.96.125134}. The last one is the surface Landau damping due to the decay of the plasma into the electron-hole pairs, which leads to an additional absorption source of the EMF in the metal \cite{PhysRevB.94.235431,Li2013}. The light-matter interactions modified by the QSEs can be effectively treated by introducing the Feibelman $d$-parameters $d_{\perp }=\int_{-\infty }^{\infty }\mathrm{d}zz\rho_\text{ind} (z)/\int_{-\infty }^{\infty}\mathrm{d}z\rho_\text{ind} (z)$ and $d_{\parallel }=\int_{-\infty }^{\infty }\mathrm{d}zz\partial_{z}K_\text{ind}^{x}(z)/\int_{-\infty }^{\infty }\mathrm{d}z\partial _{z}K_\text{ind}^{x}(z)$, with the complex-valued $\rho_\text{ind} (z)$ and $K_\text{ind}^{x}(z)$ denoting the induced charge and current density near the interface by the QSEs \cite{Goncalves2020,PhysRevLett.118.157402}, see Appendix \ref{Appen-sec1}. The real parts of $d_{\perp/\parallel}$ characterize the centroids of the induced charge and the normal derivative of the tangential current [see Fig. \ref{Fig1}(b)], while the imaginary parts measure the surface-enabled damping \cite{Stamatopoulou:22,mortensen2014generalized}. Within the formalism, the charge density is not rigorously cut off in the interface, but instead follows a Friedel oscillation and quasi-exponential decay inside and outside the metal, respectively. This leads to a correction to the classical boundary conditions and a modified response function. With the QSEs being encoded into $d_{\perp/\parallel}$, Feibelman's treatment permits us to localize the dielectric function in the bulk of the metal still as the classical Drude model $\varepsilon _{m}({\bf r},{\bf r}',\omega )\simeq\varepsilon_m(\omega)=1-\omega _{p}^{2}/[\omega(\omega +i\gamma _{p})]$ \cite{Stamatopoulou:22,PhysRevB.97.165423}, where $\omega _{p}$ is the bulk plasma frequency and $\gamma _{p}$ is the damping factor of the EMF in the metal.

\section{Dissipation suppression induced by QSEs}\label{part2}
It is easy to verify that the total excitation number $\hat{\mathcal{N}}=\sum_{i}\hat{\sigma}_{i}^{\dag}\hat{\sigma}_{i}+\int d^3{\bf r}\int d\omega \mathbf{\hat{f}}^{\dag }(\mathbf{r},\omega )\cdot \mathbf{\hat{f}}(\mathbf{r},\omega )$ is conserved in the system. Thus, if only the first QE is excited and the EMF is in the vacuum state $|\{0_{\mathbf{r},\omega }\}\rangle $ initially, then the evolved state of the total system can be expanded as $|\Psi (t)\rangle =[\sum_{i}a_{i}(t)\hat{\sigma}_{i}^{\dag }+\int \mathrm{d}^{3}\mathbf{r}\int \mathrm{d}\omega b_{\mathbf{r},\omega }(t)\mathbf{\hat{f}}^{\dag }(\mathbf{r},\omega )]|G;\{0_{\mathbf{r},\omega }\}\rangle$, where $|G\rangle $ denotes that all the QEs are in the ground state and $a_{i}(t)$ is the excited-state probability amplitude of the $i$th QE. According to the Schr\"{o}dinger equation, we have
\begin{equation}
\dot{\mathbf{a}}(t)+i\omega _{0}\mathbf{a}(t)+\int_{0}^{t}\mathrm{d}\tau\int_{0}^{\infty }\mathrm{d}\omega e^{-i\omega (t-\tau )}\mathbf{J}(\omega )\mathbf{a}(\tau )=0,  \label{integro-differential}
\end{equation}
where $\mathbf{a}(t)=(a_{1}(t),\cdot\cdot\cdot,a_{N}(t))^{T}$ is a column vector and $\mathbf{J}(\omega)$ is a matrix of $N$-body spectral density with elements $J_{ij}(\omega) =\omega ^{2}\pmb{\mu} _{i}\cdot \textrm{Im}[\mathbf{G}(\mathbf{r}_{i},\mathbf{r}_{j},\omega )]\cdot \pmb{\mu} _{j}^{\ast }/(\pi\hbar\varepsilon_{0}c^2)$ characterizing the correlations between the $i$th and $j$th QEs. Equation \eqref{integro-differential} indicates that indirect couplings among the QEs are induced by their direct interactions with the common SPP. It is easy to verify $J_{ij}(\omega)=J_{ji}(\omega)\equiv J_{|i-j|}(\omega)$. The convolution in Eq. \eqref{integro-differential} renders the dynamics non-Markovian, which is significant in quantum plasmonics due to the enhanced light-matter interactions by the sub-diffraction feature of the SPP \cite{PhysRevLett.121.227401,PhysRevA.109.033518,PhysRevB.105.245411}.

Choosing the dipole moments of the QEs being identical and normal to the interface, i.e., $\pmb{\mu}_{i}=\mu \mathbf{e}_{z}$, only the component $G_{zz}(\mathbf{r}_{i},\mathbf{r}_{j},\omega )$ contributes to the dynamics. Solving the Maxwell-Helmholtz equation under the boundary conditions modified by the QSEs, see Appendix \ref{Appen-Green}, we obtain
\begin{equation}\label{Gzz}
G_{zz}(\mathbf{r}_{i},\mathbf{r}_{j},\omega )=\int_{0}^{\infty }\frac{{\rm d}k_{s}}{4\pi }\dfrac{i\mathcal{J}_{0}(k_{s}r_{\parallel }^{ij})}{k_{d}^{2}k_{z_{d}}k_{s}^{-3}}(e^{ik_{z_{d}}z_{ij}^{-}}+r^{\text{p}}e^{ik_{z_{d}}z_{ij}^{+}}).
\end{equation}
Here $r_{\parallel}^{ij}=\sqrt{(x_{i}-x_{j})^{2}+(y_{i}-y_{j})^{2}}$ and $z_{ij}^{\pm}=z_{i}\pm z_{j}$. $\mathcal{J}_{0}$ is the zeroth-order Bessel function of the first kind. $k_s$ and $k_{z_{d/m}}$ satisfying $k_{s}^{2}+k^2_{z_{d/m}}=k_{d/m}^2$, with $k_{d/m}=\sqrt{\varepsilon _{d/m}}\omega/c$ and $\varepsilon_d$ being the permittivity of the dielectric, are the wave-vector components parallel and perpendicular to the dielectric-metal interface, respectively. The optical response of the interface to the EMF modified by the QSEs is collected in the reflection coefficient \cite{Yang2019,Goncalves2020}, see Appendix \ref{Appen-sec1},
\begin{equation}
r^{\text{p}}=\frac{\varepsilon _{m}k_{z_{d}}-\varepsilon_{d}k_{z_{m}}+i(\varepsilon _{m}-\varepsilon _{d})(k_{s}^{2}d_{\perp}-k_{z_{d}}k_{z_{m}}d_{\parallel })}{\varepsilon _{m}k_{z_{d}}+\varepsilon_{d}k_{z_{m}}-i(\varepsilon _{m}-\varepsilon _{d})(k_{s}^{2}d_{\perp}+k_{z_{d}}k_{z_{m}}d_{\parallel })}.\label{reflc}
\end{equation}
The dispersion equation of the SPP is determined by the poles of $r^{\text{p}}$, i.e.,
\begin{equation}
\frac{\varepsilon _{m}}{k_{z_{m}}}+\frac{\varepsilon _{d}}{k_{z_{d}}}%
-i(\varepsilon _{m}-\varepsilon _{d})(\frac{k_{s}^{2}d_{\perp }}{%
k_{z_{d}}k_{z_{m}}}+d_{\parallel })=0,
\label{dispersion}
\end{equation}
where $k_{s}$ is complex, and the dispersion relation for SPP can be obtained by taking the real part. In the limit of the large QE-interface distance, $d_{\perp}=d_{\parallel}=0$ and Eqs. \eqref{reflc} and \eqref{dispersion} reduce naturally to the LRA results \cite{Novotny2012,1994Dyadic}. In the weak-coupling limit, the Markovian approximate solution of Eq. \eqref{integro-differential} is $\mathbf{a}_\text{MA}(t)=\exp[-(\bar{\pmb\gamma}/2+i\bar{\pmb\omega})t]\mathbf{a}(0)$, where $\bar{\pmb\gamma}=2\pi {\bf J}(\omega_0)$ and $\bar{\pmb \omega}=\omega_0{\bf I}+\mathcal{P}\int_{0}^{\infty }{\rm d}\omega \mathbf{J}(\omega)/(\omega-\omega_0)$, with $\mathcal{P}$ being the Cauchy principal value. The positivity of $\bar{\pmb \gamma}$ results in an exponential decay of $|{\bf a}_\text{MA}(t)|$, which means that all the QEs tend to their ground state in the long-time limit \cite{PhysRevB.82.075427,PhysRevB.82.115334}. It is harmful for applying the SPP in quantum technologies.

Although the non-Markovian solution of Eq. (\ref{integro-differential}) is obtainable only by numerical calculations, its asymptotic form can be derived as follows. A Laplace transform converts Eq. (\ref{integro-differential}) into $\tilde{\mathbf{a}}(s)=[s+i\omega _{0}+\int_{0}^{\infty }\mathrm{d}\omega\frac{\mathbf{J}(\omega )}{s+i\omega } ]^{-1}\mathbf{a}(0)$. Then, $\mathbf{a}(t)$ is the inverse Laplace transform of $\tilde{\mathbf{a}}(s)$, which needs to find its poles by the equation $Y_{j}(\varpi)=\varpi$, where $\varpi =is$, $Y_{j}(\varpi)\equiv\omega _{0}-\int_{0}^{\infty }\mathrm{d}\omega \frac{A_{j}(\omega )}{\omega -\varpi }$, and $\mathbf{A}(\omega)=\mathbf{C}^{-1}\mathbf{J}(\omega )\mathbf{C}=\text{diag}[A_{1}(\omega ),...,A_{N}(\omega )]$ is the Jordan canonical form of $\mathbf{J}(\omega)$. It is interesting to find that the roots $\varpi$ multiplied by $\hbar$ are the eigenenergies of Eq. \eqref{Hami}. To prove this, we expand the eigenstate of Eq. \eqref{Hami} in the single-excitation subspace as $|\Phi \rangle =[\sum_{i=1}^{N}x_{i}\hat{\sigma}_{i}^{\dag }+\int {\rm d}^{3}\mathbf{r}\int {\rm d}\omega y_{\mathbf{r},\omega }\mathbf{\hat{f}}^{\dag }(\mathbf{r},\omega )]|G;\{0_{\mathbf{r},\omega }\}\rangle $. From $\hat{H}|\Phi \rangle =E|\Phi\rangle $, with $E$ being the eigenenergy, we have
\begin{eqnarray}
(\hbar \omega _{0}-E)x_{i} &=&\int {\rm d}\omega \int {\rm d}^{3}\mathbf{r}\frac{i\omega^{2}}{c^{2}}\sqrt{\text{Im}[\varepsilon_{m}(\omega )]\over\pi \varepsilon _{0}/\hbar}{\pmb\mu }_{i}\cdot\mathbf{G}(\mathbf{r}_{i},\mathbf{r},\omega )y_{\mathbf{r},\omega },  \label{B1} \\
(E-\hbar \omega _{0})y_{\mathbf{r},\omega } &=&\frac{i\omega ^{2}}{c^{2}}\sqrt{\text{Im}[\varepsilon _{m}(\omega )]\over\pi \varepsilon _{0}/\hbar}\sum_{i=0}^{N}{\pmb\mu }_{i}^{\ast }\cdot\mathbf{G}^{\ast }(\mathbf{r}_{i},\mathbf{r},\omega )x_{i}.  \label{B2}
\end{eqnarray}
Substituting Eq. \eqref{B2} into Eq. \eqref{B1} and using $\int {\rm d}^{3}\mathbf{r}\frac{\omega ^{2}}{c^{2}}\text{Im}[\varepsilon _{m}(\omega )]\mathbf{G}(\mathbf{r}_{i},\mathbf{r},\omega )\mathbf{G}^{\ast }(\mathbf{r}_{j},\mathbf{r},\omega )$$=\text{Im}[\mathbf{G}(\mathbf{r}_{i},\mathbf{r}_{j},\omega )]$, we obtain
$\lbrack E-\hbar \omega _{0}-\hbar ^{2}\int {\rm d}\omega \frac{\mathbf{J}(\omega )}{E-\hbar \omega }]\mathbf{x}=0$, where $\mathbf{x}=(x_{1},\cdots,x_{N})^{T}$. It has nontrivial solution only when the determinant of the coefficient matrix is zero, i.e.,
\begin{equation}
\omega _{0}-\int {\rm d}\omega \frac{A_{j}(\omega )}{\omega -E/\hbar}=\frac{E}{\hbar }.  \label{eigenenergy}
\end{equation}%
Equation \eqref{eigenenergy} has the same form as the pole equation determining the dynamics. It indicates that the dynamics of QEs characterized by $\mathbf{a}(t)$ is intrinsically determined by the features of the energy spectrum of Eq. \eqref{Hami}.

Because $Y_{j}(\varpi)$ is a decreasing function in the region of $\varpi<0$, the pole equation has a discrete root $\varpi_j^{b}$ provided $Y_{j}(0)<0$. The eigenstate of this discrete eigenenergy $\hbar\varpi^{b}_j$ falling in the band-gap regime of the SPP is called a bound state, whose formation has profound impacts on the dynamics \cite{PhysRevResearch.1.023027,PhysRevA.109.033518}. It is expected that, depending on the system parameters, $N$ bound states could be formed at most for our $N$-QE configuration. In the region of $\varpi>0$, $Y_{j}(\varpi)$ is ill-defined and the pole equation has an infinite number of roots, which form a continuous energy band. Using the Cauchy's residue theorem, we obtain
\begin{equation}
\mathbf{a}(t)=\mathbf{Z}(t)+\int_{0}^{\infty }\frac{\mathrm{d}\varpi }{2\pi }[\tilde{\mathbf{a}}(0^{+}-i\varpi )-\tilde{\mathbf{a}}(0^{-}-i\varpi )]e^{-i\varpi t},
\end{equation}
where $\mathbf{Z}(t)=\sum_{j=1}^{M}\text{Res}[\tilde{\mathbf{a}}(-i\varpi_{j}^{b})]e^{-i\varpi _{j}^{b}t}$, with $\text{Res}[\cdot]$ denoting the residue contributed by the bound states, and $M$ being the number of the bound states. The second term is from a branch cut formed by the continuous energy band and tends to zero in long-time limit due to the out-of-phase interference in continuously changing frequencies. Thus, if the bound state is absent, then $\lim_{t\rightarrow\infty}\mathbf{a}(t)=0$ characterizes a complete dissipation. It is consistent with the Markovian approximate \cite{PhysRevB.82.115334,PhysRevB.89.041402} and the previous results \cite{Goncalves2020,PhysRevB.104.L201405}. If $M$ bound states are formed, then $\lim_{t\rightarrow\infty}\mathbf{a}(t)=\mathbf{Z}(t)$ implies a suppression of dissipation. Being absent in the previous results, it indicates that the non-Markovian effect plays a constructive role in suppressing the dissipation of the QEs caused by the lossy SPP.

\section{Numerical results}\label{numr}
It was previously found that the QSEs exert a reversible energy exchange between QEs and SPP in the strong-coupling regime \cite{Goncalves2020,PhysRevB.104.L201405}. However, the QEs exclusively dissipate to their ground states in the long-time limit. Our above analysis shows that the effects are also beneficial to suppress the QE dissipation and to preserve the coherent energy exchange among the QEs mediated by the SPP. To verify this, we choose sodium as the metal, because the QSEs, e.g., the electron spill-out effects, are larger than the ones in conventional noble metal \cite{PhysRevLett.118.157402,Babaze2023} and the $d$-parameters can be obtained by ab initio calculations, while the ones in the noble metals require a more demanding atomistic treatment \cite{gonccalves2023interrogating} . By numerically fitting the data extracted from ab initio TDDFT via a sum of Lorentzians \cite{Goncalves2020,PhysRevB.104.L201405}, we obtain $d_{\perp}$, see Fig. \ref{Fig1}(c). For this charge-neutral interface, $d_{\parallel}$ vanishes \cite{liebsch1997electronic}.

\begin{figure}[tbp]
\centering
\includegraphics[width=0.8\columnwidth]{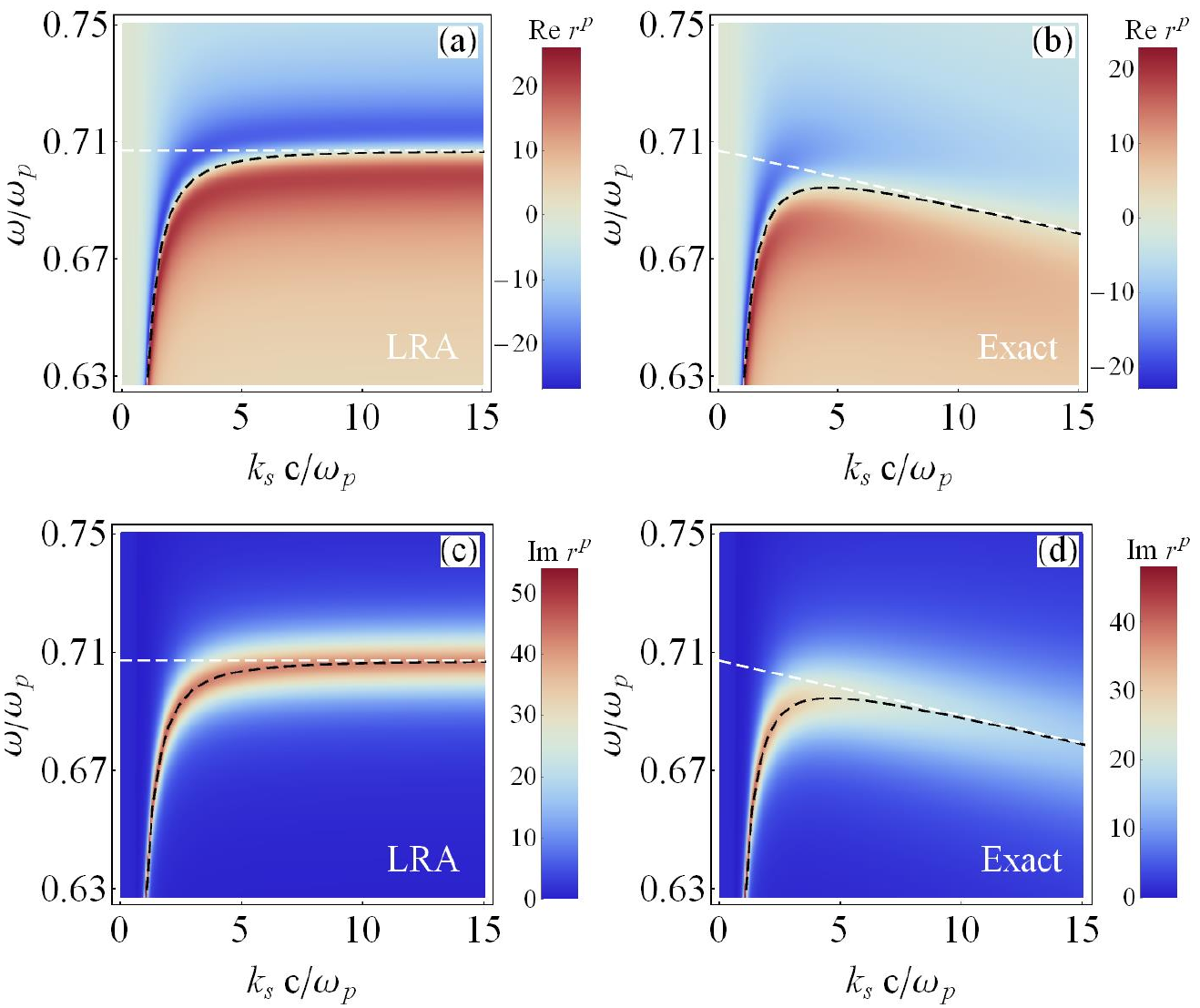}
\caption{Real parts of $r^{\text{p}}$ under the (a) LRA and (b) exact treatments. Imaginary parts of $r^{\text{p}}$ under the (c) LRA and (d) exact treatments. The black dashed lines are the dispersion relation of SPP and the white dashed lines are the corresponding relations of $\omega _{sp}$. We use $\hbar\gamma_p=0.1$ eV and $\hbar\omega_p=5.9$ eV.}\label{Fig2}
\end{figure}
With $d_{\perp}$ at hand, we can calculate the reflection coefficient $r^{\text{p}}$ under both the exact and LRA treatments, see Fig. \ref{Fig2}. Their difference reveals the QSEs of the interface to the optical response of the EMF at the nanoscale. A common character of the exact and LRA results is that the real part of $r^\text{p}$ equals to zero, see Figs. \ref{Fig2}(a) and \ref{Fig2}(b), and the imaginary part shows a peak, see Figs. \ref{Fig2}(c) and \ref{Fig2}(d), at the line of the dispersion relation of the SPP denoted by the black dashed lines. In the region of small wavevector $k_{s}$, Figs. \ref{Fig2}(a)-(d) illustrate the negligible influence of QSEs on the dispersion relations of the SPP. However, as $k_{s}$ increases, the reflection coefficient $r^{\text{p}}$ in the exact result shows significant differences from the one under the LRA. According to Eq. \eqref{dispersion}, neglecting the imaginary part, the dielectric function of the metal can be expressed as $\varepsilon _{m}(\omega )=1-\omega _{p}^{2}/\omega ^{2}$, from which we obtain the dispersion relation of SPP shown with black dashed line. When $k_{s}\gg k_{0}$, $k_{z_{d/m}}\rightarrow ik_{s}$, then Eq. \eqref{dispersion} possesses the asymptotic solution of $\omega _{sp}=\frac{\omega _{p}}{\sqrt{2}}\sqrt{1-k_{s}d_{\perp}}$. Under the LRA, $d_{\perp}=0$, the black dispersion curves converge to the position of the white dashed lines $\omega _{sp}=\omega _{cl}\equiv \frac{\omega _{p}}{\sqrt{2}}$ in Figs. \ref{Fig2}(a) and \ref{Fig2}(c). Figures \ref{Fig2}(b) and \ref{Fig2}(d) show the exact dispersion curves converge to the position of the white dashed lines $\omega _{sp}=\frac{\omega _{p}}{\sqrt{2}}\sqrt{1-k_{s}d_{\perp}}$. 
Compared to results in dispersion-relation measurement under the LRA, the resonance redshift and the plasmonic dispersion is broadened, which is a reflection of electron spill-out and Landau damping, respectively \cite{PhysRevLett.74.1558,PhysRevLett.110.263901,Mandal2013}.

\begin{figure}[tbp]
\centering
\includegraphics[width=0.8\columnwidth]{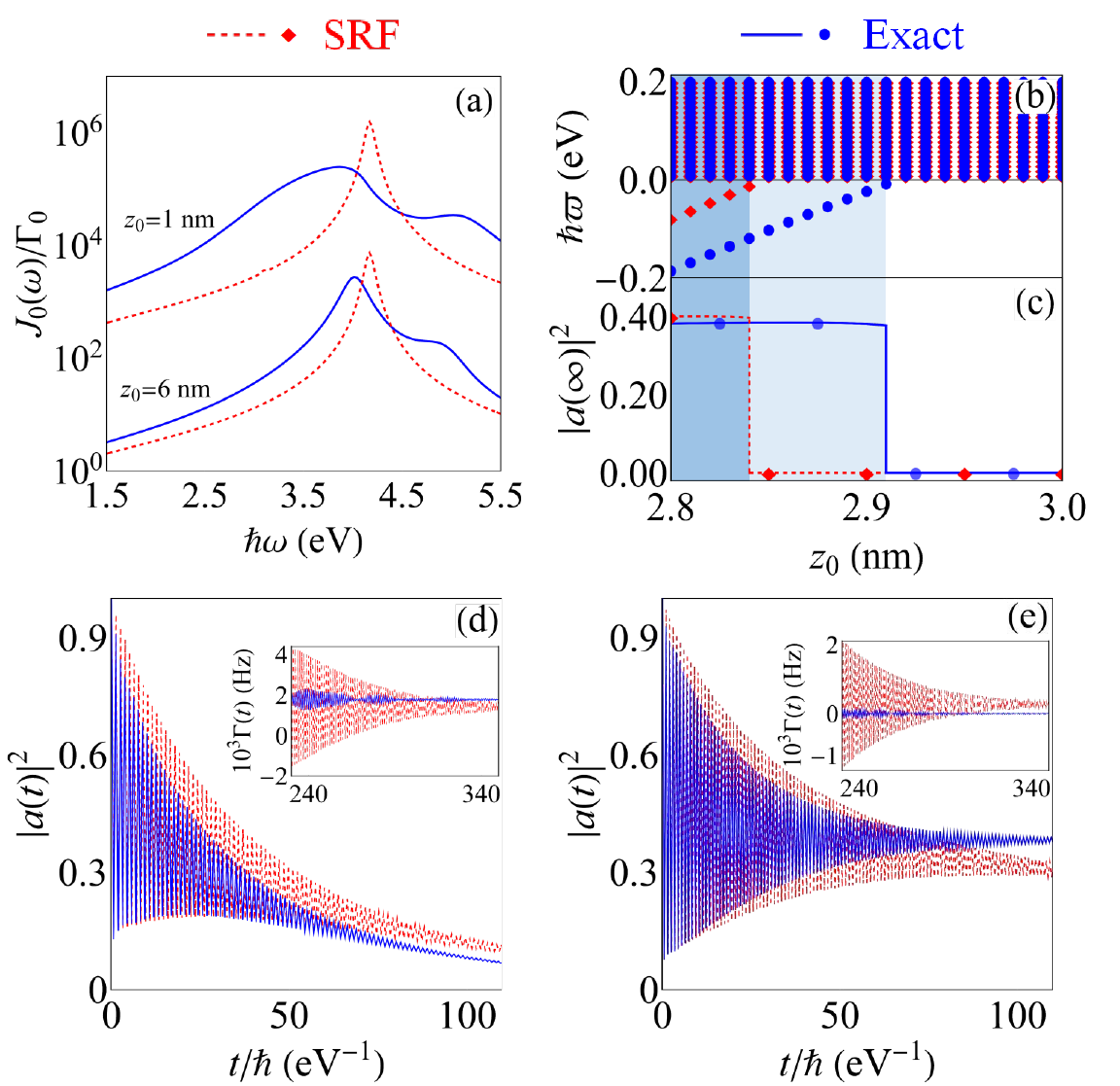}
\caption{(a) Ratio of $J_{0}({\omega })$ and free-space spontaneous emission rate $\Gamma _{0}=\omega _{0}^{3}\mu ^{2}/3\pi \hbar \varepsilon _{0}c^{3}$, (b) energy spectrum by solving Eq. \eqref{eigenenergy}, and (c) $|a(\infty)|^2$  by solving Eq. \eqref{integro-differential} in the presence (blue dots) and absence (red diamonds) of the QSEs in different $z_0$ when $t/\hbar=10^3$ eV$^{-1}$. The dark blue and light blue shades in (b)(c) highlight the areas with the formation of bound state and its constructive role in dissipation suppression. The blue solid and red dashed lines in (c) are the analytic form $L_0^2$ in the presence and absence of the QSEs. Evolution of $|a(t)|^2$ when (d) $z_0=3.5$ nm and (e) $2.9$ nm. The insets of (d) and (e) show the decay rate $\Gamma(t)\equiv-\text{Re}[\dot{a}(t)/a(t)]$. We use $\hbar \omega _{0}=2.3$ eV and $N=1$. Other parameters are the same as Fig. \ref{Fig2}.}\label{Fig3}
\end{figure}

The substitution of $r^\text{p}$ into Eq. \eqref{Gzz} results in $G_{zz}$ and thus the spectral density $J_0(\omega)$, which characterizes the coupling strength of light-matter interaction mediated by SPP in the presence of the QSEs. Figure \ref{Fig3}(a) shows $J_{0}(\omega)$ in different $z_0$. Under the LRA,  $J_0(\omega)$ shows one peak, which enables the SPP to be effectively described by a pseudo cavity mode with a Lorentzian spectrum. This method has been widely used in studying the strong QE-SPP coupling \cite{PhysRevB.108.L180409,PhysRevB.89.041402}. In contrast to the LRA result in spectroscopy measurement, the QSEs cause a red shift and a clear shoulder in the high-frequency regime, which relates to the electron spill-out \cite{Goncalves2020,PhysRevB.104.L201405}, and a broadening of the resonance plasmon peak on $J_0(\omega)$, which is attributed to the Landau damping \cite{Khurgin2017}. These features make the Lorentzian fitting to $J_0(\omega)$ in the pseudo-cavity method insufficient. The smaller the QE-surface distance, the stronger the QSEs. A common character between the exact and LRA $J_0(\omega)$ is their enhancement over the spontaneous emission rate in free space in five orders of magnitude when $z_0$ reaches the nanoscale. It is due to the sub-diffraction confinement to the EMF by the SPP. Manifesting a strong QE-SPP coupling, such an enhanced $J_0(\omega)$ makes the SPP an ideal platform in exploring strong light-matter interactions. However, the QSEs contribute additional loss sources to the SPP, which makes the lifetime of the SPP even shorter than that under the LRA. It severely restricts the applications of the SPP in quantum technologies. In the following, we focus on the influence of QSEs on the non-Markovian dynamics of the QEs.

First, we consider the case of $N=1$. The spectral density becomes $\mathbf{J}(\omega )=J_{0}(\omega)$. The Laplace transform of $a(t)$ reduces to $\tilde{a}(s)=[s+i\omega_0+\int_0^\infty{J_0(\omega)\over s+i\omega}{\rm d}\omega]^{-1}$. If a bound state with eigenenergy $\hbar\varpi^b$ is formed, then the residue contributed by it is
\begin{eqnarray}
\mathbf{Z}(t)=\lim_{s\rightarrow -i\varpi^b}(s+i\varpi^b)e^{st}\tilde{a}(s)= L_0e^{-i\varpi^b t},
\end{eqnarray}
where $L_{0}=[1+\int {\rm d}\omega \frac{J_{0}(\omega )}{(\omega-\varpi ^{b})^{2}}]^{-1}$. Then the excited-state population has a long-time limit $|a(\infty)|^2=L_0^2$. To reveal the mechanism behind the non-Markovian dynamics, we plot the energy spectrum and $|a(\infty)|^2$ in Figs. \ref{Fig3}(b) and \ref{Fig3}(c). Figure \ref{Fig3}(b) is energy spectrum that consists of a continuous band and one branches of bound state in the band gap, respectively. Unlike the complete vanishing in the Markovian approximation, $|a(\infty)|^2$ matching with our analytic $L_0^2$ exhibits an excited-state preservation once the bound state is formed. It is surprising that the QE can preserve its excited-state population even when it interacts with the SPP experiencing severe loss in such an absorptive medium. The governed mechanism is just the bound state, which is valid irrespective of whether the LRA is made or not. Compared with the LRA result, the QE-surface distance $z_0$ in supporting the bound state is enlarged in the exact result when the QSEs are considered. It indicates that the QSEs are helpful in forming the bound state and suppressing the dissipation of the QE. The rapid oscillations in the evolution of $|a(t)|^2$ in Figs. \ref{Fig3}(d) and \ref{Fig3}(e) manifest the non-Markovian back-action effect. In the case of a large $z_0$, the bound state is absent for both the exact and LRA results and thus no qualitative difference is found between them, see Fig. \ref{Fig3}(d). It indicates the validity of the LRA in the regime of a large QE-interface separation. With decreasing $z_0$, the bound state is still absent for the LRA one and the QE keeps decaying to the ground state, but it is present for the exact result, which results in a preserved excited-state population, see Fig. \ref{Fig3}(e). Thus, the QSEs signify their actions on the QE not only in the transient dynamics, but also in the steady state. This is further confirmed by that the decay rate $\Gamma(t)\equiv-\text{Re}[\dot{a}(t)/a(t)]$ tends to a positive value in the absence of the bound state, meaning a complete dissipation, while tends to zero in the presence of the bound state, meaning a halt of the dissipation, see the inset of Figs. \ref{Fig3}(d) and \ref{Fig3}(e).

An intuitive understanding on these bound-state favored behaviors are as follows. Consider that $M$ bound states are formed. The evolved state $|\Psi(t)\rangle$ of an initial state $|\Psi(0)\rangle$ of the QE-SPP system can be expanded in the complete basis formed by the eigenstates of $\hat{H}$, i.e., $\hat{H}|\Phi_\alpha\rangle=E_\alpha|\Phi_\alpha\rangle$, as
\begin{equation}
|\Psi(t)\rangle=\sum_{j=1}^{M}c^{b}_{j}e^{-iE^{b}_{j}t/\hbar}|\Phi_{j}^{b}\rangle+\sum_{\alpha\in\text{Band}}c_{\alpha} e^{-iE_{\alpha} t/\hbar}|\Phi_{\alpha}\rangle,\label{dvd1}
\end{equation}
where $c^{b}_{j}=\langle\Phi_{j}^{b}|\Psi(0)\rangle$ and $c_{\alpha}=\langle\Phi_{\alpha}|\Psi(0)\rangle$ are the components of the $j$th bound state and $\alpha$th energy-band eigenstate in $|\Psi(0)\rangle$. The excited-state population of the $l$th QE, i.e., $P_l(t)=\langle\Psi(t)|\hat{\sigma}_l^\dag\hat{\sigma}_l|\Psi(t)\rangle$, is calculated to be
\begin{eqnarray}
P_l(t)&=&\sum_{j,j'=1}^M c^{b*}_{j'}c^{b}_je^{-i(E_j^b-E_{j'}^b)t\over\hbar} x^{b*}_{j',l}x^b_{j,l}+\sum_{\alpha,\alpha'\in \text{Band}} c^{*}_{\alpha'} c_\alpha e^{-i(E_\alpha-E_{\alpha'})t\over\hbar}  x^{*}_{\alpha',l}x_{\alpha,l}\nonumber\\
&&+\sum_{j=1}^M\sum_{\alpha\in \text{Band}}c^{*}_{\alpha}c^{b}_je^{-i(E_j^b-E_{\alpha})t\over\hbar} x^{*}_{\alpha,l}x_{j,l}+\text{c.c.},
\end{eqnarray}
where $x_{j,l}^b$ and $x_{\alpha,l}$ are the excited-state probability amplitudes of the $l$th QE in the $j$th bound state and the $\alpha$ band state, respectively. The second and third terms contain the oscillating frequencies $E_{\alpha}/\hbar$, which are continuously summed in the continuous energy band. Such terms tend to zero due to the out-of-phase interference of the different components in the long-time limit. Thus, only the contributions of the bound states survive in the long-time limit. When $M=1$, the excited-state population tends to a constant nonzero value. When $M=2$, the excited-state population tends to a periodic oscillation with a frequency $|E^b_1-E^b_2|/\hbar$. This clearly demonstrates the distinguished role of the bound states in suppressing the dissipation of the QEs induced by the SPP.

\begin{figure}[tbp]
\centering
\includegraphics[width=0.8\columnwidth]{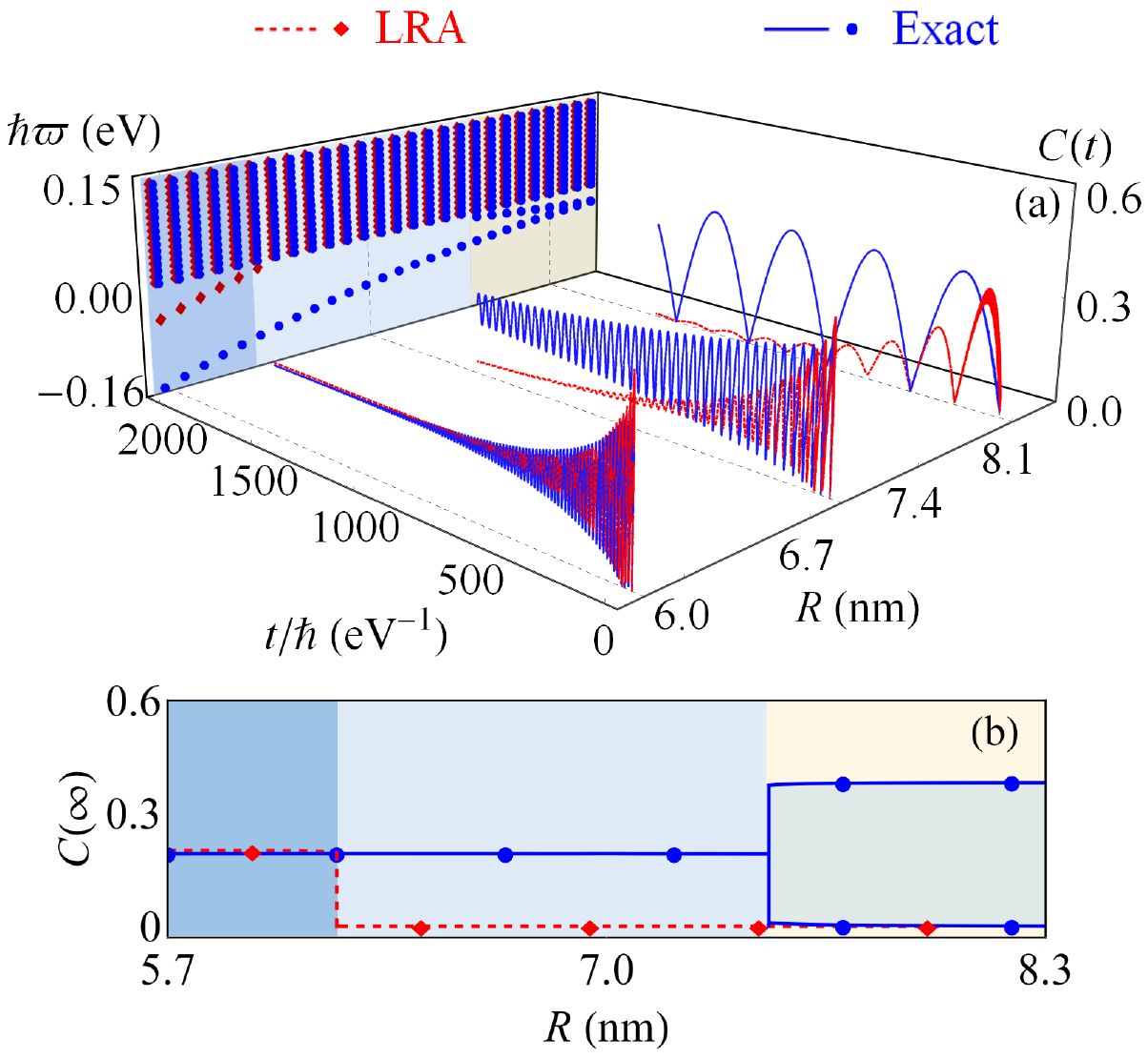}
\caption{(a) Energy spectrum by solving Eq. \eqref{eigenenergy} and evolution of $C(t)$ evaluated under the LRA and exact calculations in different $R$. (b) Long-time values of $C(t)$ obtained by solving Eq. \eqref{integro-differential} (solid and dashed lines) and bound-state analysis (dots and diamonds) in Eq. (\ref{case-2}). The colored shades highlight the areas with the formation of bound state and its constructive role in generating persistent entanglement between QEs. We use $z_0=2.9$ nm and $N=2$. Other parameters are the same as Fig. \ref{Fig2}.} \label{Fig4}
\end{figure}

Second, we consider $N=2$. The spectral-density matrix is
$\mathbf{J}(\omega )=\left(
\begin{array}{cc}
J_{0}(\omega ) & J_{1}(\omega ) \\
J_{1}(\omega ) & J_{0}(\omega )
\end{array}
\right) $ and the initial condition is $\mathbf{a}(0)=(1,0)^{T}$. The Laplace transform of ${\bf a}(t)$ is
$\tilde{\bf a}(s)=\left(
\begin{array}{cc}
1 & 1 \\
1 & -1
\end{array}
\right) \left(
\begin{array}{c}
l_{01}^{(1)}(s) \\
l_{01}^{(2)}(s)
\end{array}%
\right)$, where $l_{mn}^{(j)}(s)=[s+i\omega _{0}+\int {\rm d}\omega \frac{J_{m}(\omega )-(-1)^{j}J_{n}(\omega )}{s+i\omega  }]^{-1}/2$. The residue contributed by the $l$th bound state with eigenenergy $\hbar\varpi_l^b$ evaluated by $Y_l(\varpi)=\varpi$ to the inverse Laplace transform of $l_{01}^{(j)}(s)$ is $\lim_{s\rightarrow -i\varpi_l^b}(s+i\varpi_l^b)e^{st}l_{01}^{(j)}(s)=L_{01}^{(j)} e^{-i\varpi_l^b t} \delta_{l,j}$, where $L_{01}^{(j)}(\varpi )=[\partial _sl_{01}^{(j)}|_{s=-i\varpi _{j}^{b}}]^{-1}$. Therefore, if $M$ bound states are formed, then we have
\begin{equation}
\mathbf{Z}(t)=\sum_{j=1}^{M}L_{01}^{(j)}\left(
\begin{array}{c}
1 \\
(-1)^{j+1}%
\end{array}%
\right) e^{-i\varpi ^{b}_{j}t}.
\end{equation} We study the entanglement of the QEs mediated by the SPP. The entanglement is measured by concurrence $C=\max {\{0,\sqrt{\lambda _{1}}-\sqrt{\lambda _{2}}-\sqrt{\lambda _{3}}-\sqrt{\lambda _{4}}\}}$, where $\lambda _{i}$ are the eigenvalues of $\rho (\hat{\sigma}_{y}\otimes \hat{\sigma}_{y})\rho ^{\ast }(\hat{\sigma}_{y}\otimes \hat{\sigma}_{y})$ in decreasing order and $\rho$ is the reduced density matrix of the QEs \cite{PhysRevLett.80.2245}. We can derive
\begin{equation}\label{case-2}
\lim_{t\rightarrow \infty }C(t)=\left\{
\begin{array}{cc}
0, ~~~~~~~~~~~~~~~~~~~~~~~~~~~& M=0 \\
2L_{01}^{(1)2}, ~~~~~~~~~~~~~~~~~~~~~& M=1 \\
2|L_{01}^{(1)2}-L_{01}^{(2)2}+L(t)|, & M=2,\end{array}
\right.
\end{equation}
where $L(t)=2iL_{01}^{(1)}L_{01}^{(2)}\sin [(\varpi _{1}^{b}-\varpi _{2}^{b})t]$. It is found that, if the bound state is absent, then no entanglement can be established in the long-time limit, which is consistent with the Markovian approximate result. If one bound state is present, then a stable entanglement is generated. If two bound states are present, then a persistently oscillating entanglement is created, which signifies a lossless energy exchange between the QEs mediated by the SPP. These expectations are verified by the numerical calculation, see Figs. \ref{Fig4}(a) and \ref{Fig4}(b). We find from the energy spectrum in Fig. \ref{Fig4}(a) that, in contrast to the LRA result that only one bound state is present, two bound states are present when the QE separation $R$ is large in our exact result. The lossless energy exchange favored by the two bound states in the presence of the QSEs makes the two QEs coherently and permanently correlated in a manner of Rabi-like oscillation. This implies that our system could act as a quantum bus to remotely interconnect the QEs as nodes of a quantum network \cite{ZHOU20191,doi:10.1126/science.aax3766,PhysRevLett.124.213601,Zheng2023}. It is out of one's expectation that the SPP experiencing a severe loss due to the QSEs still can mediate a coherent and permanent long-range correlation between the QEs.

\section{Discussion and conclusion}\label{conclusions}
Our scheme is realizable in the state-of-the-art experiments. The detection of QSEs and its significant influence on plasmon-mediated light-matter interactions in nanostructured systems have been realized \cite{Yang2019,Boroviks2022,10.1063/1.1951057,doi:10.1126/sciadv.adn5227,Chen2025,Ye2023}. The QEs could be CdSe quantum dot or fluorescent molecule (e.g. the J-aggregate), which can be selectively prepared in a single-excited state by optical or electrical drivings and have been widely used in experiments of quantum plasmonics \cite{PhysRevB.80.155307,FIDDER1990529,PhysRevLett.111.026804}. We choose sodium as the metal, whose QSEs are larger than those in conventional noble metals \cite{PhysRevLett.118.157402,Babaze2023}. Modern manufacturing methods, including the nanopositioning techniques, the scanning QE methods, and the microfluidic flow control, enable the precise adjustment on positions of the QEs, thereby providing a strong support to achieve the strong QE-SPP coupling \cite{PhysRevLett.118.126802,PhysRevLett.93.036404,PhysRevLett.103.053602}. The bound state and its pivotal role in governing the non-Markovian dynamics of QEs have been observed in circuit QED \cite{Yanbing2017,PhysRevX.9.011021}, cold-atom systems \cite{Krinner2018,Kwon2022}, and topological systems \cite{PhysRevX.11.011015}, which lays a firm foundation for the feasibility of implementing our scheme in experiments.

In summary, we have investigated the QSEs on the QE-SPP interactions in a planar dielectric-metal nanostructure. We have discovered a mechanism to overcome the dissipation of the QEs caused by the lossy SPP in the presence of the additional metal absorption introduced by the surface-assisted nonlocal optical response, electron spill-out, and Landau damping. It is revealed that, with the formation of one and two bound states of the QE-SPP system, the QEs, with the dissipation being efficiently suppressed, would be permanently correlated to a stable value and coherently correlated in a manner of Rabi-like oscillation in the long-time condition, respectively. In contrast to the results under the classical LRA \cite{PhysRevResearch.1.023027,PhysRevA.109.033518}, our result highlights a constructive role played by the QSEs in establishing the long-distance coherent correlations. It is out of one's general belief that the QSE-enhanced loss of the SPP is harmful for it to mediate a long-distance entanglement between QEs. The result enriches our understanding on plasmon-mediated light-matter interactions at the nanoscale and is beneficial for applying the SPP in quantum network.

\appendix

\section{Feibelman $d$-parameter method}\label{Appen-sec1}


The Feibelman method reintroduces the electronic length scales by amending the classical boundary conditions with a set of mesoscopic complex surface response functions. They enable a leading-order-accurate incorporation of the QSEs including nonlocal response, electron spill-out, and surface-enabled Landau damping. The lowest-order nonclassical corrections to the macroscopic electromagnetic results are realized by introducing the scattering coefficients of the EMF on the metal-dielectric interface \cite{FEIBELMAN1982287,PhysRevLett.118.157402}. Consider that the metal and the dielectric have a planar interface at $z=0$. When an EMF with a potential $\phi _{\text{ext}}(\mathbf{r})=e^{ikx+kz}$ impinges from the dielectric region ($z>0$) on the interface, a quantum charge density $\rho _{\text{ind}}(\mathbf{r})=\rho _{\text{ind}}(z)e^{ikx}$ is induced on the interface. An induced potential $\phi _{\text{ind}}(\mathbf{r})=\phi _{\text{ind}}(z)e^{ikx}$ generated by $\rho _{\text{ind}}(z)$ is determined by the Coulomb's law as $\phi _{\text{ind}}(z)=\int_{z_{1}}^{z_{2}}\frac{{\rm d}z^{\prime }}{2\varepsilon _{0}k}e^{-k|z-z^{\prime }|}\rho_{\text{ind}} (z^{\prime })$, where $z_{1}$ ($<0$) and $z_{2}$ ($>0$) are the positions far from the interface where $\rho_{\text{ind}} (z)$ vanishes. Because $\rho _{\text{ind}}$ is sharply peaked near the interface, the asymptotic behavior of $\phi _{\text{ind}}(z)$ beyond $z_{1}$ and $z_{2}$ is obtainable by expending it around $kz^{\prime }=0$. A multipole expansion results in
$\phi_{\text{ind}}(z)\simeq\frac{e^{-k|z|}}{2\varepsilon _{0}k}\sigma [1+\text{sgn}(z)kd_{\perp }]$, where $\sigma =\int_{z_{1}}^{z_{2}}\rho_{\text{ind}} (z)dz$ and
\begin{equation}
d_{\bot}={1\over\sigma}\int_{z_{1}}^{z_{2}}z\rho_{\text{ind}} (z){\rm d}z \label{d1}
\end{equation}
characterizes the position of the centroid of $\rho_{\text{ind}} (z)$.

To obtain the optical scattering coefficients with non-classical corrections, we define an auxiliary potential $\phi^{\infty }(z)=\left\{
\begin{array}{cc}
e^{kz}+r^{\text{nr}}e^{-kz}, & z>0 \\
t^{\text{nr}}e^{kz}, & z<0
\end{array}\right.$, which agrees with the actual total potential $\phi _{\text{ext}}(z)+$ $\phi _{\text{ind}}(z)$ in the asymptotic regions $|z|\geq |z_{1/2}|$. Here, $r^{\text{nr}}$\ and $t^{\text{nr}}$ are the reflection and transmission coefficients in the non-retarded region. According to $\phi_{\text{ind}}(z)$, we obtain $r^{\text{nr}}=\frac{1}{2\varepsilon _{0}k}\sigma (1+kd_{\perp })$ and $t^{\text{nr}}=1+\frac{1}{2\varepsilon _{0}k}\sigma
(1-kd_{\perp })$. The continuity condition of $\phi^{\infty }(z)$ and its derivative at $z=0$ results in
\begin{equation}
(1-t^{\text{nr}})(1+kd_{\bot })+r^{\text{nr}}(1-kd_{\bot })=0,  \label{con1}
\end{equation}
which is the first effective boundary condition. A current $\mathbf{K}_{\text{ind}}(\mathbf{r})$ is also induced by the EMF near the interface. Its impact on the scattering coefficients is viewed as the second effective boundary condition. To obtain it, a set of auxiliary classical charge and current densities, i.e., $\rho ^{\infty }(\mathbf{r})$ and $\mathbf{K}^{\infty }(\mathbf{r})$, interrelated by the continuity equation ${\pmb\nabla}\cdot \mathbf{K}^{\infty}(\mathbf{r})-i\omega\rho ^{\infty }(\mathbf{r})=0$ has to be introduced. Here, $\mathbf{K}^{\infty }(\mathbf{r})=\mathbf{K}^{\infty }(z)e^{ikx}$ approaches $\mathbf{K}_{\text{ind}}(\mathbf{r})$ in the asymptotic regions $|z|\geq |z_{1,2}|$ and $\rho ^{\infty }(\mathbf{r})=\sigma ^{\infty }\delta (z)e^{ikx}$, with $\sigma ^{\infty }$ being the asymptotic surface charge density. By integrating the continuity equation crossing the interface and exploiting the asymptotic equality between $\mathbf{K}^{\infty }$ and $\mathbf{K}_{\text{ind}}$, we obtain the second boundary condition
\begin{equation}
\varepsilon _{d}(1-r^{\text{nr}})+(\varepsilon _{d}-1)kd_{\parallel }(1+r^{\text{nr}})=[\varepsilon _{m}(1+kd_{\parallel })-kd_{\parallel }]t^{\text{nr}},  \label{con2}
\end{equation}
where $d_{\parallel }=\frac{\int_{z_{1}}^{z_{2}}[K_{\text{ind}}^{x}(z)-K_{x}^{\infty }(z)]{\rm d}z}{K_{x}^{\infty }(0^{-})-K_{x}^{\infty }(0^{+})}$.
In the limit of $k\ll |z_{1,2}^{-1}|$, the asymptotic current $K_{x}^{\infty }(z)$ changes slowly over the region $z\in \lbrack z_{1},z_{2}]$. It allows for
\begin{eqnarray}
d_{\parallel }&\simeq& 
\frac{\int_{z_{1}}^{z_{2}}z\partial _{z}K_{\text{ind}}^{x}(z){\rm d}z}{\int_{z_{1}}^{z_{2}}\partial _{z}K_{\text{ind}}^{x}(z){\rm d}z}.\label{smdpx}
\end{eqnarray}
Furthermore, because the field is localized in the region near the interface, one can safely extend the integration domain in Eqs. \eqref{d1} and \eqref{smdpx} to infinity. Solving the effective boundary conditions in Eqs. \eqref{con1} and \eqref{con2}, the modified scattering coefficients can be obtained. 

The Feilbelman $d$-parameters can be interpreted in terms of effective boundary quantities at the interface. $d_{\perp }$ is expressed in terms of a surface dipole density $\pmb{\pi } =-\varepsilon _{0}d_{\perp }[[\nabla _{\perp }\phi ]]\mathbf{\hat{n}}$, with $\nabla _{\perp }=\hat{\bf n}\cdot{\pmb\nabla} _{\perp}$, while $d_{\parallel}$ is incorporated as a current density $\mathbf{K} =-i\varepsilon _{0}\omega d_{\parallel }[[\varepsilon {\pmb \nabla}_{\parallel }\phi ]]$ \cite{PhysRevLett.118.157402,Yang2019}. Here, $[[\cdot ]]$ denotes the discontinuities, i.e. $[[\nabla _{\perp }\phi]]=\nabla _{\perp }\phi (\mathbf{r}_{\parallel }^{+})-\nabla _{\perp }\phi (\mathbf{r}_{\parallel }^{-})$ and $[[\varepsilon {\pmb\nabla}_{\parallel }\phi]]=\varepsilon ^{+}{\pmb\nabla}_{\parallel }\phi (\mathbf{r}_{\parallel}^{+})-\varepsilon^{-}{\pmb \nabla}_{\parallel }\phi (\mathbf{r}_{\parallel}^{-})$, where the bulk permittivities $\varepsilon ^{+}\equiv \varepsilon _{d}$ outside (i.e. at $\mathbf{r}_{\parallel }^{+}=\mathbf{r}_{\parallel }+0^{+}\mathbf{\hat{n}}$) and $\varepsilon ^{-}\equiv \varepsilon _{m}$ inside (i.e. at $\mathbf{r}_{\parallel }^{-}=\mathbf{r}_{\parallel }-0^{+}\mathbf{\hat{n}}$) the interface extended along $\mathbf{r}_{\parallel }=(x,y)$, and $\mathbf{\hat{n}}$ is the outward vector normal to the interface. According to $-{\pmb \nabla} \phi \rightarrow \mathbf{E}$ and $-\varepsilon _{0}\varepsilon {\pmb\nabla} \phi \rightarrow \mathbf{D}$, we obtain their forms in the retarded regime as $\pmb{\pi }=\varepsilon _{0}d_{\perp }[\mathbf{\hat{n}}\cdot (\mathbf{E}_{d}-\mathbf{E}_{m})]\mathbf{\hat{n}}$ and $\mathbf{K} =i\omega d_{\parallel }[\mathbf{\hat{n}}\times (\mathbf{D}_{d}-\mathbf{D}_{m})\times \mathbf{\hat{n}}]$ \cite{Yang2019,Goncalves2020}. Incorporating these surface densities in the macroscopic boundary conditions $[[\mathbf{E}_{\parallel }]]=-\varepsilon _{0}^{-1}{\pmb\nabla }_{\parallel }({\pmb\pi}\cdot\hat{\bf n})$ and $[[\mathbf{H}_{\parallel}]]=\mathbf{K}\times \mathbf{\hat{n}}$ \cite{Novotny2012}, we obtain the modified boundary conditions as
\begin{eqnarray}
\mathbf{E}_{d\parallel }-\mathbf{E}_{m\parallel } &=&-d_{\perp }{\pmb\nabla }_{\parallel }[E_{d\perp }-E_{m\perp }],  \label{bou-con1} \\
\mathbf{H}_{d\parallel }-\mathbf{H}_{m\parallel } &=&i\omega d_{\parallel }[\mathbf{\hat{n}}\times (\mathbf{D}_{d}-\mathbf{D}_{m})\times \mathbf{\hat{n}]}. \label{bou-con2}
\end{eqnarray}
They indicate that the QSEs make the parallel components of the electric and magnetic fields discontinuous, which is different from the ones in the classical theory.

When the QE-interface separation is small, the transverse magnetic (TM) waves dominates the plasmon-enhanced light-matter interactions. Hence, we seek the scattering coefficients of the TM waves with a $p$ polarization. Their wave functions are
\begin{eqnarray}
\mathbf{H}&=&\left\{\begin{array}{cc}
\left( e^{-k_{z_d}z}+r^{\text{p}}e^{ik_{z_d}z}\right) e^{i\left(k_{s}x-\omega t\right) }\mathbf{\hat{y}},&z>0 \\
t^{\text{p}}e^{-ik_{z_m}z}e^{i\left( k_{s}x-\omega t\right) }\mathbf{\hat{y}},& z<0
\end{array}\right.,   \label{H1}\\
\mathbf{E}&=&\left\{\begin{array}{cc}
\big[ E_{x_d}(z) \mathbf{\hat{x}}+E_{z_d}(z)\mathbf{\hat{z}}\big] e^{i( k_{s}x-\omega t) }, & z>0 \\
\big[ E_{x_m}(z) \mathbf{\hat{x}}+E_{z_m}(z)\mathbf{\hat{z}}\big] e^{i( k_{s}x-\omega t) },& z<0
\end{array}\right., ~~~~~~~\label{E1}
\end{eqnarray}
where $r^{\text{p}}$ and $t^{\text{p}}$ are the modified reflection and transmission coefficients, $k_{s}$\ and $k_{z_{d/m}}$ are the wave-vector components parallel and perpendicular to the interface, satisfying $k_{s}^{2}+k_{z_{d/m}}^{2}=k_{d/m}^2=\varepsilon _{d/m}\omega ^{2}/c^{2}$. Substituting Eqs. \eqref{H1} and \eqref{E1} into Eqs. \eqref{bou-con1} and \eqref{bou-con2} and retaining only the linear-order terms of $k_{s}d_{\perp ,\parallel }$, we obtain
\begin{eqnarray}
r^{\text{p}} =\frac{\varepsilon _{m}k_{z_{d}}-\varepsilon_{d}k_{z_{m}}+i(\varepsilon _{m}-\varepsilon _{d})(k_{s}^{2}d_{\perp}-k_{z_{d}}k_{z_{m}}d_{\parallel })}{\varepsilon _{m}k_{z_{d}}+\varepsilon_{d}k_{z_{m}}-i(\varepsilon _{m}-\varepsilon _{d})(k_{s}^{2}d_{\perp
}+k_{z_{d}}k_{z_{m}}d_{\parallel })}, ~~~~~ \label{reflectionrr} \\
t^{\text{p}} =\frac{2\varepsilon _{d}k_{z_{m}}}{\varepsilon_{m}k_{z_{d}}+\varepsilon _{d}k_{z_{m}}-i(\varepsilon _{m}-\varepsilon_{d})(k_{s}^{2}d_{\perp }+k_{z_{d}}k_{z_{m}}d_{\parallel })}.~~~~~\label{reflectiontt}
\end{eqnarray}
The optical response of the interface to the EMF is collected in the reflection and transmission coefficients. The dispersion relation of the SPP is determined by the real part of the poles, i.e.,
\begin{equation}
\frac{\varepsilon _{m}}{k_{z_{m}}}+\frac{\varepsilon _{d}}{k_{z_{d}}}%
-i(\varepsilon _{m}-\varepsilon _{d})(\frac{k_{s}^{2}d_{\perp }}{%
k_{z_{d}}k_{z_{m}}}+d_{\parallel })=0.\label{dfdsap}
\end{equation}
In the classical limit of the large QE-interface distance, Eqs. \eqref{reflectionrr}, \eqref{reflectiontt}, and \eqref{dfdsap} return to the results in classical macroscopic electrodynamics \cite{Novotny2012,1994Dyadic}.

Note that the applicability of the Feibelman method is restricted to typical sizes $\geq 1$ nm \cite{Yang2019}. In truly microscopic scales $< 1$ nm, other quantum mechanical effects, such as electron tunneling and quantum size effects, become important, and a fully microscopic approach has to be developed in the treatment of quantum plasmonics. Furthermore, the Feibelman method originally describes properties of a planar interface. For interfaces with finite curvature, it requires the assumption of ``local flatness". Thus, for arbitrary shape of nanostructures, the Feibelman method can be practically implemented with principal curvatures $|\mathcal{K}_{i}|\ll |d_{\bot ,\Vert }|$ \cite{Yang2019}. However, the conclusion in our work is general for arbitrary shapes of the nanostructures, as long as the Feibelman method is applicable.

\section{Green's tensor}\label{Appen-Green}
For a layered structure composed by a metal in the regime of $z<0$ and a dielectric in the regime $z>0$ and a source point positioned above the interface, the Green's tensor is given by \cite{Novotny2012}
\begin{equation*}
\mathbf{G}(\mathbf{r},\mathbf{r}^{\prime},\omega )=\left\{
\begin{array}{cc}
\mathbf{G}_{0}(\mathbf{r},\mathbf{r}^{\prime},\omega )+\mathbf{G}_{R}(\mathbf{r},\mathbf{r}^{\prime},\omega ), & z>0\\
\mathbf{G}_{T}(\mathbf{r},\mathbf{r}^{\prime},\omega ), & z<0
\end{array}\right.
\end{equation*}
where $\mathbf{G}_{0/R/T}(\mathbf{r},\mathbf{r}^{\prime},\omega )$ are the Green's tensors contributed by the free space, the reflected, and the transmitted field, respectively. Their explicit forms are
\begin{eqnarray}
\mathbf{G}_{0}(\mathbf{r},\mathbf{r}^{\prime },\omega ) &=&\int\frac{i{\rm d}^2\mathbf{k}_{\parallel }}{8\pi^{2}}\dfrac{e^{ik_{z_{d}}(z-z^{\prime })}}{k_{d}^{2}k_{z_{d}}}\left(
\begin{array}{ccc}
k_{d}^{2}-k_{x}^{2} & -k_{x}k_{y} & \mp k_{x}k_{z_{d}} \\
-k_{x}k_{y} & k_{d}^{2}-k_{y}^{2} & \mp k_{y}k_{z_{d}} \\
\mp k_{x}k_{z_{d}} & \mp k_{y}k_{z_{d}} & k_{d}^{2}-k_{z_{d}}^{2}\end{array}\right) , \\
\mathbf{G}_{R}(\mathbf{r},\mathbf{r}^{\prime },\omega ) &=&\int\frac{{\rm d}^2\mathbf{k}_{\parallel }}{8i\pi^{2}}\frac{r^{\text{p}}e^{ik_{z_{d}}(z+z^{\prime })}}{k_{d}^{2}(k_{x}^{2}+k_{y}^{2})}\left(
\begin{array}{ccc}
k_{x}^{2}k_{z_{d}} & k_{x}k_{y}k_{z_{d}} & k_{x}(k_{x}^{2}+k_{y}^{2}) \\
k_{x}k_{y}k_{z_{d}} & k_{y}^{2}k_{z_{d}} & k_{y}(k_{x}^{2}+k_{y}^{2}) \\
-k_{x}(k_{x}^{2}+k_{y}^{2}) & -k_{y}(k_{x}^{2}+k_{y}^{2}) &
{(k_{x}^{2}+k_{y}^{2})^{2}\over-k_{z_{d}}}\end{array}
\right) , ~~\\
\mathbf{G}_{T}(\mathbf{r},\mathbf{r}^{\prime },\omega ) &=&\int\frac{i{\rm d}^2\mathbf{k}_{\parallel }}{8\pi^{2}}\frac{t^{\text{p}}e^{i(k_{z_{d}}z^{\prime }-k_{z_{m}}z)}}{k_{d}k_{m}(k_{x}^{2}+k_{y}^{2})}\left(\begin{array}{ccc}
k_{x}^{2}k_{z_{m}} & k_{x}k_{y}k_{z_{d}} &
{k_{x}(k_{x}^{2}+k_{y}^{2})k_{z_{m}}\over k_{z_{d}}} \\
k_{x}k_{y}k_{z_{d}} & k_{y}^{2}k_{z_{d}} &
{k_{y}(k_{x}^{2}+k_{y}^{2})k_{z_{m}}\over k_{z_{d}}} \\
k_{x}(k_{x}^{2}+k_{y}^{2}) & k_{y}(k_{x}^{2}+k_{y}^{2}) &
{(k_{x}^{2}+k_{y}^{2})^{2}\over k_{z_{d}}}\end{array}
\right) ,~~~~~~
\end{eqnarray}
where ${\rm d}^2\mathbf{k}_{\parallel }=e^{i[k_{x}(x-x^{\prime})+k_{y}(y-y^{\prime })]}{\rm d}k_{x}{\rm d}k_{y}$, the upper sign applies for $z>z^{\prime }$, and the lower sign applies for $z<z^{\prime }$.

Choosing the dipole moment perpendicular to the interface, i.e., $\pmb{\mu}_{i}=\mu \mathbf{e}_{z}$, only the $zz$ component of the Green's tensor contributes to the QE-SPP interactions in the regime $z>0$.
Substituting $k_{x}=k_{s}\cos \varphi$ and $k_{y}=k_{s}\sin \varphi$ and integrating $\varphi$, we obtain
\begin{equation*}
G_{zz}(\mathbf{r},\mathbf{r}^{\prime },\omega )=\int_{0}^{\infty }\frac{{\rm d}k_{s}}{4\pi }\dfrac{i\mathcal{J}_{0}(k_{s}r_{\parallel })}{k_{d}^{2}k_{z_{d}}k_{s}^{-3}} (e^{ik_{z_{d}}z^-}+r^{\text{p}}e^{ik_{z_{d}}z^+}),
\end{equation*}
where $k_{s}^{2}=k_{x}^{2}+k_{y}^{2}=k_{d}^{2}-k_{z_{d}}^{2}$, $r_{\parallel}=\sqrt{(x-x^{\prime })^{2}+(y-y^{\prime })^2}$, $z^\pm=z\pm z'$, and $\mathcal{J}_{0}(x)$ is the zeroth-order Bessel function of the first kind.

\section*{Acknowledgment}
This work is supported by the National Natural Science Foundation of China (Grants No. 12275109, No. 12074106, and No. 12247101), the Innovation Program for Quantum Science and Technology of China (Grant No. 2023ZD0300904), and the Natural Science Foundation of Henan Province (Grant No. 242300421165).

\section*{Disclosures} The authors declare no conflicts of interest.


\bibliography{SRF}

\end{document}